\documentclass{caosp307}
\usepackage{graphicx}
\usepackage{epstopdf}
\usepackage{natbib}
\articleNo{XXX}
\pubyear{2020}
\volume{50}
\volnumber{3}
\firstpage{672}
\received{March 24, 2020}
\accepted{April 15, 2020}

\def\BibTeX{{\rm B\kern-.05em{\sc i\kern-.025em b}\kern-.08em
             T\kern-.1667em\lower.7ex\hbox{E}\kern-.125emX}}

\begin{document}
\hauthor{V.\,Andreoli,U.\,Munari}
\title{LAMOST J202629.80+423652.0 is not a symbiotic star}

\author{
         V.\,Andreoli \inst{1} 
      \and 
        U.\,Munari \inst{2}   
       }
\institute{
          ANS Collaboration, Astronomical Observatory, 36012 Asiago (VI), Italy
          \and
          INAF Padova Astronomical Observatory, I-36012 Asiago (VI), Italy
          }
      
\date{XXX.X.XXXX}

\maketitle

\begin{abstract}
LAMOST J202629.80+423652.0 has been recently classified as a new symbiotic
star containing a long-period Mira, surrounded by dust (D-type) and
displaying in the optical spectra high ionization emission lines, including
the Raman-scattered OVI at 6825 \AA.  We have observed LAMOST
J202629.80+423652.0 photometrically in the BVRI bands and spectroscopically
over the 3500-8000~\AA\ range.  We have found it to be a normal G8\,IV
sub-giant star, deprived of any emission line in its spectrum, and reddened
by $E_{B-V}$=0.35 mag.  Combining our photometry with data from all-sky
patrol surveys, we find LAMOST J202629.80+423652.0 to be non variable, so not
pulsating as a Mira.  We have compiled from existing sources its spectral
energy distribution, extending well into the mid-Infrared, and found it
completely dominated by the G8\,IV photospheric stellar emission, without any 
sign of circumstellar dust. We therefore conclude that 
LAMOST J202629.80+423652.0 is not a symbiotic star, nor it is pulsating or been
enshrouded in dust.
\end{abstract}

\section{Introduction}
\label{intr}

Symbiotic stars (SySt) are binary systems composed by a red giant star (RG)
that transfers material to a degenerate companion which can be either a
white dwarf (WD) or a neutron star (NS).  SySt are broadly divided into two
major groups (see the recent review by Munari 2019):
those {\it accreting-only} whose optical spectra are dominated by the RG
with no or weak emission lines, and the {\it burning-type} displaying a
strong nebular continuum and a rich emission line spectrum: they originate in
the wind of the RG largely ionized by the very hot and luminous WD companion
that is nuclearly burning at the surface in stable conditions.
The accreting-only SySt usually requires satellite
UV/X-ray observations to be discovered (cf.  the prototype SU Lyn, Mukai et
al.  2016), while the burning StSy can easily be discovered through the
whole Galaxy and the Local Group thanks to the prominent emission lines.
In about 20\% of the known SySt (see recent catalogs of symbiotic stars by
Belczy{\'n}ski et al.  2000, and Akras et al.  2019), the cool giant
pulsates as a Mira and this usually comes also with the presence of warm
circumstellar dust ($\sim$500$-$1000 K), leading to a classification
as D-type according to the scheme introduced by Allen (1982).

Li et al. (2015) have announced the discovery of two new SySt during the
LAMOST all-sky spectroscopic survey.  LAMOST is a 4m Schmidt telescope
located in Xinglong (China), feeding light to 16 spectrographs via 4000
optical fibers.  The spectrum presented by Li et al.  for one of these two
new SySt, namely LAMOST J202629.80+423652.0 (hereafter LM), looks weird and
we suspected to be the result of problematic sky subtraction, so decided to
investigate the matter by acquiring new spectra and supplementing them with
acquisition of BVRI photometry.  We report in this paper the results of
these observations.

\section{Observations}

\subsection{Spectroscopy}

Low resolution spectra of LM have been obtained with the 1.22m telescope
operated in Asiago by the University of Padova.  A Boller \& Chivens spectrograph is
attached to the Cassegrain focus feeding light to an ANDOR iDus DU44OA
camera, that houses a back-illuminated E2V 42-100 CCD (2048 X 512 array,
13.5 $\mu$m pixel).  A 300 ln mm$^{-1}$ grating blazed at 5000 \AA\ was
adopted, allowing to cover the 3300-8000 \AA\ range at 2.31 \AA/pix
dispersion.  Data reduction was carried out in IRAF and involved the usual
steps on bias and dark removal, flat-fielding, variance-weighted spectrum
tracing, sky subtraction, wavelength calibration via FeHeAr lamp, heliocentric correction and
flux calibration (via observations of the nearby spectrophotometric standard
HR 7867). LM has been observed twice on Dec 6 and 14, 2019, 
exposing for 16min on both occasions. The spectra look identical and their
average is plotted in Figure~1. A comparison is there provided with
HD~188512, a template for the G8\,IV spectral class according to Yamashita
et al. (1977), observed with exactly the same telescope setup as for LM,
and selected from the Asiago Spectral Database (U. Munari, in preparation).

\begin{figure}
	\centerline{
		\includegraphics[width=12.cm,clip=]{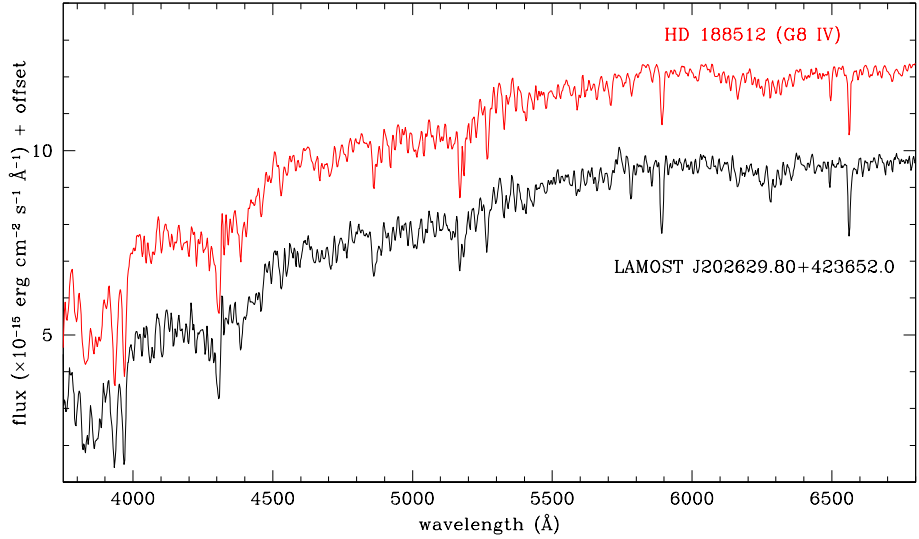}}
	\caption{Our spectrum of LAMOST J202629.80+423652.0 compared 
        to that of HD~188512 (+offset), template for the G8\,IV spectral 
	type.}
	\label{f1}
\end{figure}

\subsection{Photometry}

Optical photometry of LM in the Landolt (1992) photometric system has been obtained with a
Richey-Chretien 0.4m f/8 telescope located in Monte Baldo (Verona, Italy),
equipped with a Moravian G4 CCD camera (36$\times$36mm active area, 12$\mu$m
size) and B,V,R,I photometric filters from Astrodon. The B filter is of the
new version, corrected for the red-leak of its original version (see Munari and Moretti
2012). The photometry of LM that we have collected in five distinct nights
is presented in Table~1.

Data reduction has involved all the usual steps for bias, dark, flat with
calibration images collected during the same observing nights. We adopted
aperture photometry because the sparse field around LM did not required
PSF-fitting procedures. The transformation from the local to the Landolt
standard system was carried out via color equations calibrated on a
photometric sequence recorded on the same frames as LM:
\begin{eqnarray}
\nonumber
V   &=& v + \alpha_v \times (v-i) + \gamma_v \\  \nonumber
B-V &=& \beta_{bv} \times (b-v) + \delta_{bv} \\ \nonumber
V-R &=& \beta_{vr} \times (v-r) + \delta_{vr} \\ \nonumber
V-I &=& \beta_{vi} \times (v-i) + \delta_{vi}    \nonumber
\end{eqnarray}
The local photometric sequence has been extracted from APASS DR8 survey 
(Henden and Munari 2014), ported to the Landolt system via the
transformations calibrated by Munari et al.
(2014).  The errors quoted in Table~1 are the quadratic sum
of the Poissonian and the transformation contributions.

\begin{table}
	\small
	\begin{center}
		\caption{Our BVRI photometry (on the Landolt system of
			equatorial standards) of LAMOST J202629.80+423652.0.}
		\label{t1}
		\begin{tabular}{@{}c@{~~~}c@{~~}c@{~~}c@{~~}c@{}}
			\hline
			obs. date & B & V & R & \multicolumn{1}{c}{I} \\
			\hline
			2019-12-07.8957 &   & 14.439 $\pm$ 0.009 & 13.766 $\pm$ 0.010 & 13.041 $\pm$ 0.014 \\ 
			2019-12-14.8155 &   & 14.456 $\pm$ 0.006 & 13.773 $\pm$ 0.007 & 13.119 $\pm$ 0.016 \\
			2020-01-08.8245 & 15.629 $\pm$ 0.032 & 14.454 $\pm$ 0.011 & 13.760 $\pm$ 0.010 & 13.100 $\pm$ 0.015 \\
			2020-01-09.8242 & 15.603 $\pm$ 0.061 & 14.478 $\pm$ 0.029 & 13.737 $\pm$ 0.017 & 13.076 $\pm$ 0.025 \\
			2020-01-16.7712 & 15.580 $\pm$ 0.019 & 14.454 $\pm$ 0.006 &   &   \\
			&&&&\\
			weighted mean & 15.610 $\pm$ 0.020 & 14.463 $\pm$ 0.004 & 13.755 $\pm$ 0.005 & 13.084 $\pm$ 0.008\\
			\hline \hline
		\end{tabular}
	\end{center}
\end{table}

\section{Results}
\subsection{A normal spectral appearance}

The spectrum of LM presented in Figure~1 is that of a normal, sub-giant star
of the G8\,IV spectral type, abundant in the Solar
Neighborhood as indicated by Gaia DR2 stellar census (Gaia
Collaboration 2018).  No Mira is known to have such an early spectral type
(cf.  VSX database\footnote{https://www.aavso.org/vsx/}).
In our spectrum of LM no emission line is visible.

Li et al.  (2015) report the presence of strong H$\alpha$, H$\beta$ and [OIII]
emission lines in their spectrum of LM.  They come actually from the
sky-background emission originating from the huge HII region IC~1318
associated to nearby $\gamma$ Cyg (10$^\circ$ angular extent).  In Figure~2
we present the sky-background at 15arcsec away from LM as recorded on our
spectrum.  It looks identical to the spectrum attributed to LM by Li et al.  (2015).

Li et al.  (2015) also report about the detection in their spectrum of LM of
Raman scattered OVI lines at 6830 and 7088 \AA\ (Schmid 1989).  These are actually normal
OH lines forming in the Earth's atmosphere, frequently used for wavelength
calibration of high resolution spectra at far-red wavelengths (cf. Osterbrock
et al.  1996).  In Figure~3 we plot the far-red portion of the
sky-background at 15arcsec distance from LM taken from our spectrum of the
program star, with superimposed the identification of Meinel
rotation-vibration bands of OH.

We therefore conclude this section noting that nothing in the spectrum of LM
supports a classification as a SySt: all lines and features seen in
emission by Li et al.  (2015) seems actually due to inaccurate handling of
sky-background subtractions.

\begin{figure}
	\centerline{
		\includegraphics[width=12.3cm,clip=]{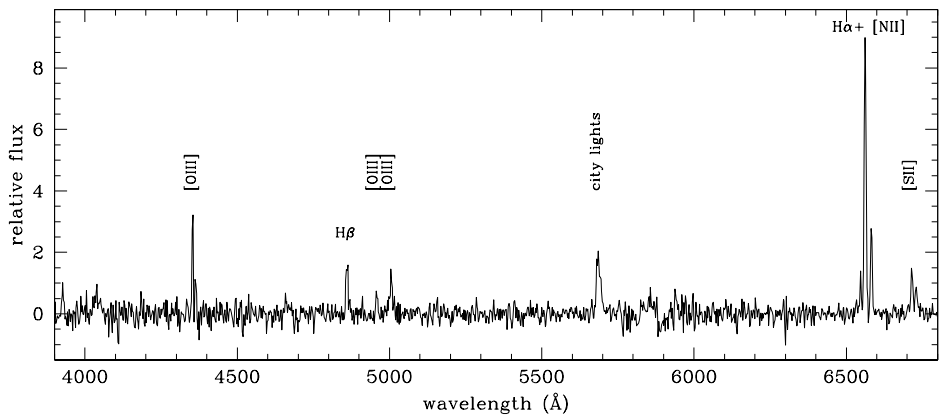}}
	\caption{Emission lines from the sky-background around
		LAMOST J202629.80+423652.0 in our spectrum, due to diffuse 
		emission from the HII region around $\gamma$~Cyg.}
	\label{f2}
\end{figure}

\begin{figure}
	\centerline{
		\includegraphics[width=12.3cm,clip=]{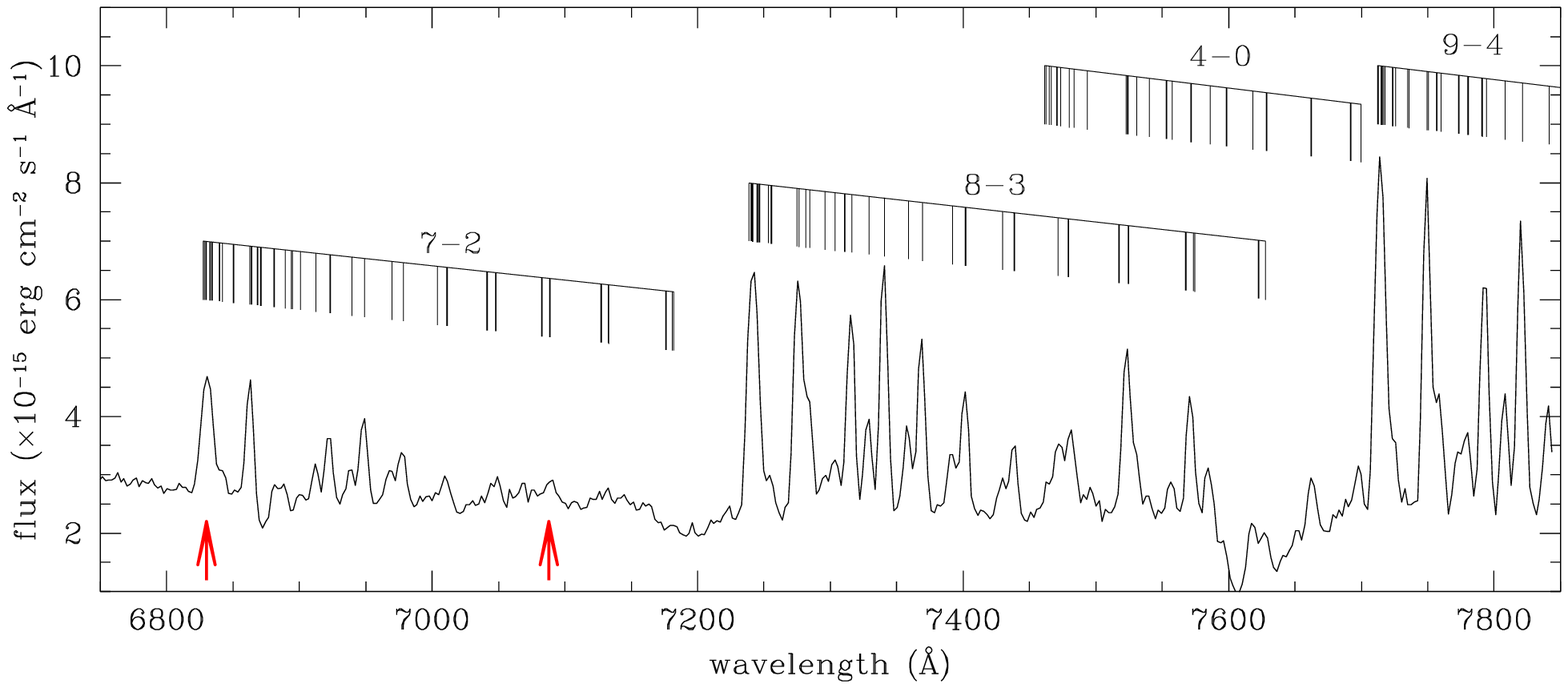}}
	\caption{Emission in the far-red of the sky-background around 
		LAMOST J202629.80+423652.0 in our spectrum,
		highlighting the Earth's OH emission (with Meinel rotation-vibration
		bands identified). The red arrows points to telluric OH lines
		labelled by Li et al. (2015) as due to Raman scattered OVI 6825,
		7088 \AA.} 
	\label{f3}
\end{figure}

\subsection{No photometric variability}

Li et al.  (2015) also classified LM as a pulsating Mira.  Our photometry in
Table~1 extends over too short a time interval to draw any conclusion about 
a possible variability affecting LM.  To investigate the matter futher, we
retrieved the patrol photometry of LM collected
by ASASSN (Shappee et al.  2014, Jayasinghe et al.  2019) and ZTF
all-sky patrol surveys (Masci et al.  2019, Bellm et al.  2019).  The ASAS-SN and ZTF
photometry in the $g^\prime$, $r^\prime$, and $V$ bands is plotted in
Figure~4.  It extends from 2015 to 2019.  No variability is present in
excess of the observational noise, and this precludes the classification
reported by Li et al.  (2015) of LM as a Mira.

\begin{figure}[thp]
	\centerline{
		\includegraphics[width=12.3 cm,clip=]{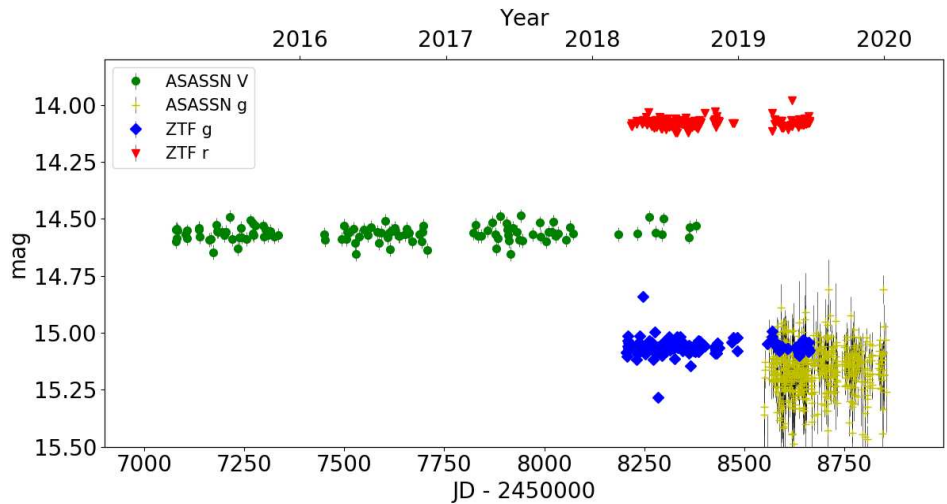}}
	\caption{Photometric data of LAMOST J202629.80+423652.0 obtained 
        by ASAS-SN and ZTF sky patrol surveys.}
	\label{f4}
\end{figure}

\subsection{No circumstellar dust}

Li et al.  (2015) attributed a D-type classification to LM, i.e.  the presence
of warm dust around it.  As illustrated by Allen (1982) the excess emission
from such warm dust (500$-$1000 K) is already evident in the $H$
band (1.6 $\mu$m) and become prominent in the $K$ band (2.2 $\mu$m).  To
investigate the spectral energy distribution (SED) of LM, we take advantage
of the absence of photometric variability (established in the previous
section) that allows to combine non-simultaneous data gathered by different surveys
over the whole wavelength range.  The SED for LM built from our BVRI
observations (Table~1), 2MASS JHK$_s$ and AllWISE W$_{\rm 1}$W$_{\rm 2}$W$_{\rm
3}$ data is presented in Figure~5, together with the reference SED for a
typical G8\,IV field star using colors from Fitzgerald (1970),
Cousins (1976), and Koornneef et al.  (1983).  The two become a fine match once
a correction for $E_{B-V}$=0.35 reddening is applied to LM.
In conclusion, the SEDs in Figure~5 excludes the presence in LM of any dust hotter
that $\sim$100~$^\circ$K (constrain set by W$_{\rm 3}$).

\begin{figure}[thp]
	\centerline{
		\includegraphics[width=12.3cm,clip=]{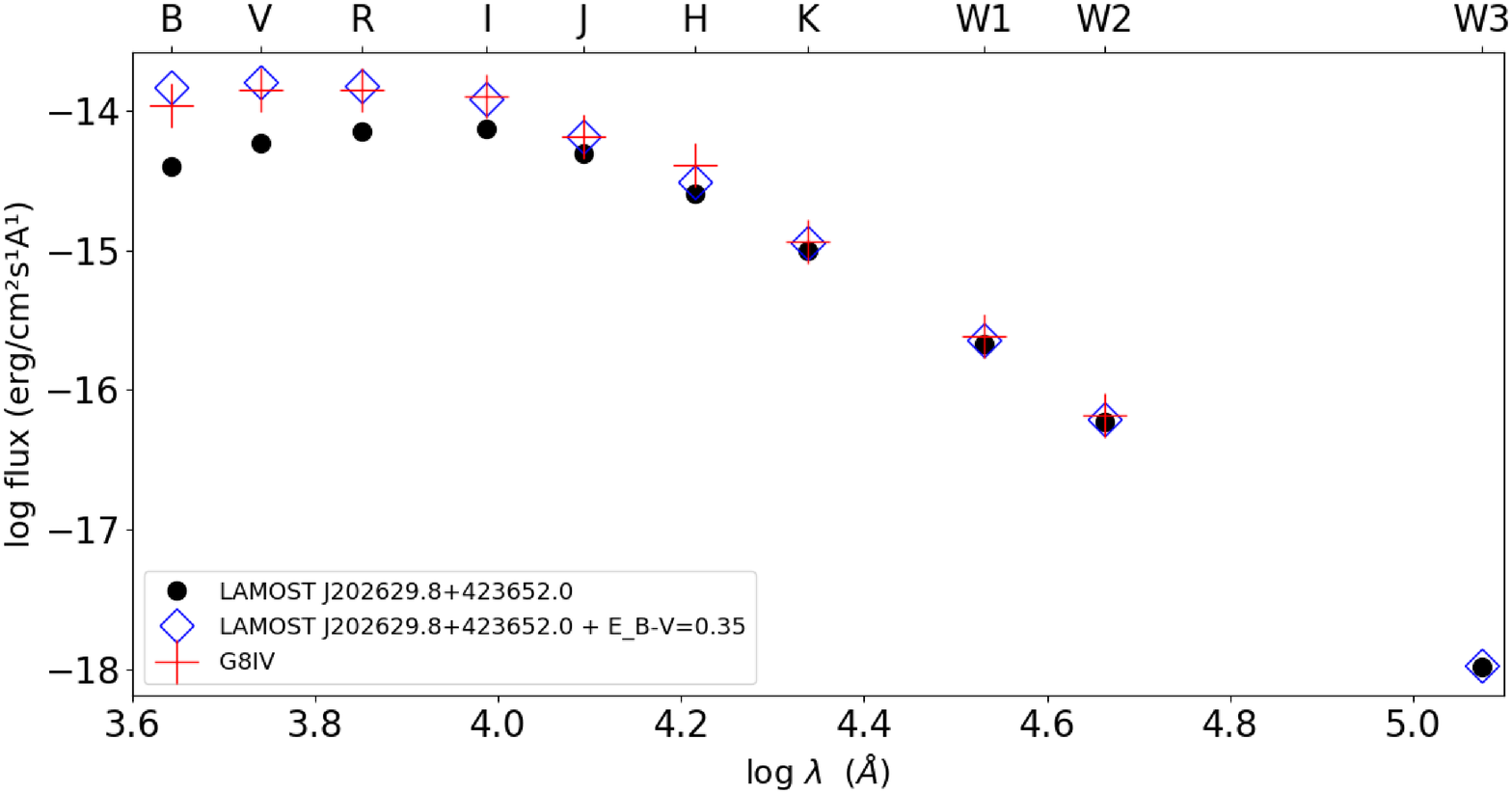}}
	\caption{Spectral energy distribution of LAMOST J202629.80+423652.0
as observed (black dots), and after correction for $E_{B-V}$=0.35 reddening
(blue diamonds). The red crosses illustrate the intrinsic energy distribution of 
a G8\,IV star from tabular data.}
	\label{f5}
\end{figure}

\subsection{Fundamental parameter}

The GAIA second release (GAIA-DR2) lists a parallax $\pi$ $= \ 0.761 \pm 0.018$
(mas) for LM. The fractional error is less than 20\% and therefore is straightforward to 
derived the distance to LM as d=1.31~kpc by direct inversion of the Gaia parallax.
At such a distance along the line of sight to LM, the Bayestar 3D Galactic extinction map (Green et al.
2019) lists a reddening of $E(g-r)=0.41$, that
transforms to  $E(B-V)_{0}=0.35$ for a standard $R_V=3.1$ interstellar reddening raw
following the relations for G-type stars by Fiorucci and Munari (2003). This is in excellent agreement with
the value of the reddening necessary to bring the SED of LM in agreement
with that of a typical G8\,IV star in Figure~5. With such a reddening and distance,
the mean $V$-mag value in Table~1 transforms to an absolute magnitude for LM of
$M(V)$=2.78. This is in fine agreement with the absolute magnitudes reported
by Sowell et al. (2007) that list M(V)=5.50, 3.20 and 1.35 for G8 stars of
luminosity class V, IV and III, respectively. Accordingt to tabular values
in Straizys (1992), such a star is characterized by a temperature of 5100~K,
a luminosity of 4.4~L$_\odot$ and a radius 2.7~R$_\odot$.

\section{Conclusions}

We have obtained new photometric and spectroscopic observations of LAMOST J202629.80+423652.0
that combined with literature data shows it to be a normal G8\,IV sub-giant star, and not a 
symbiotic star. Such a classification by Li et al.  (2015) appears due to
inaccurate sky-background subtraction to their spectra.

\bibliographystyle{caosp307}
\bibliography{LM.bib}
\end{document}